\documentclass[aps,prb,showpacs,floatfix,twocolumn]{revtex4}

\usepackage{graphicx}
\usepackage{bm}
\begin{document}
\bibliographystyle{plaint}

\preprint{APS/123-QED}

\title{The Electronic Structure of Europium Chalcogenides and Pnictides}
\author{M. Horne$^1$, P. Strange$^{1}$, W.\,M. Temmerman$^2$, Z. Szotek$^2$,
A. Svane$^3$ and H. Winter$^4$}
\affiliation{$^1$Department of Physics, Keele University, Staffordshire, ST5 5DY, England\\
$^2$Daresbury Laboratory, Warrington WA4 4AD, England\\
$3$Institute of Physics and Astronomy, University of Aarhus, DK-8000,
Aarhus, Denmark.\\
$4$INFP, Forschungzentrum Karlsruhe GmbH, Postfach 3640, D-76021, Karlsruhe,
Germany}
\date{\today}
%\linespacing{2}
%
\begin{abstract}
The electronic structure of some europium chalcogenides and pnictides is 
calculated using the {\it ab-initio} self-interaction corrected 
local-spin-density approximation (SIC-LSD). This approach allows both a 
localised description of the rare earth $f-$electrons and an itinerant 
description of $s$, $p$ and $d$-electrons. Localising different numbers of 
$f$-electrons on the rare earth atom corresponds to different nominal valencies, 
and the total energies can be compared, providing a first-principles 
description of valency. All the chalcogenides are found to be insulators in
the ferromagnetic state and to have a divalent configuration. For the pnictides 
we find that EuN is half-metallic and the rest are normal metals. However a 
valence change occurs as we go down the pnictide column of the Periodic Table. 
EuN and EuP are trivalent, EuAs is only just trivalent and EuSb is found to be 
divalent. Our results suggest that these materials may find application in 
spintronic and spin filtering devices. 
\end{abstract}
%
% **********************************************************************
\maketitle
\section{Introduction}
\label{sec:intro}
Rare earth compounds attract considerable experimental and theoretical 
attention due to the intricate electronic properties relating to the highly 
correlated $f$-electrons. In the atomic state most rare earth elements are
divalent, but in the solid state the majority form trivalent ions. 
The elemental rare earths all become trivalent with the exception of Eu and
Yb which are divalent. Europium compounds can occur in divalent and trivalent 
configurations. Europium chalcogenides and most of the pnictides crystallise 
in the simple NaCl crystal structure and so form a series that can be 
studied within first principles theory relatively easily. Recently the 
chalcogenides have attracted a lot of attention due to their potential 
applications in spintronic and spin filtering devices.

In a recent paper Steeneken {\it et al.}\cite{steen} have presented evidence 
that EuO is a ferromagnetic semiconductor where the charge carriers exhibit 
almost 100\% spin-polarisation. In this paper we perform self-consistent 
electronic structure calculations to investigate, and gain more insight into, 
this claim. The EuS system has recently been used as a spin filter in a hybrid
Gd/EuS/Al device which showed a large magnetoresistance.\cite{LeClair} 
Unfortunately the low T$_{c}$ of 16.8 K in EuS makes it unsuitable for
commercial applications, but nonetheless this does make it worthwhile to 
study, in a systematic manner, the electronic and magnetic properties of the 
remaining Eu chalcogenides.

Traditional calculations of the electronic structure of rare earth materials
treat the $s,p$ and $d$ electrons as itinerant, while the $f$ electrons are
treated as atomic-like. However, treating electrons within the same material
using two different theories is less than satisfactory. Recently
Self-Interaction Corrections (SIC)\cite{ZP} to the Local Spin Density (LSD) 
approximation to Density Functional Theory (DFT) \cite{hohen65,Kohn+Vash} 
have provided us with a successful method of treating all the electrons in 
rare earth materials on an equal footing. The self-interaction correction 
represents an $f$-electron localization energy, and the question of which 
$f^n$ configuration of the rare earth ion will be the most stable is a 
competition between this localization energy and the energy which an 
electron may gain by hybridizing into the conduction band states. This energy 
balance is a very delicate quantity and may be changed easily by altering 
external parameters.

In this paper we briefly describe the SIC-LSD formalism in section II 
and go on to discuss our calculation of the electronic structure of the Eu 
chalcogenides and pnictides in both the divalent and trivalent states in 
section III. We conclude the paper in section IV.

\section{The SIC-LSD formalism}

The standard method for making {\it ab-initio} calculations of the
properties of materials is density functional theory with a local
approximation for the exchange-correlation energy. A drawback of 
this approach is that it introduces a spurious self-interaction for each 
electron. While these are usually negligible they become important when
localisation phenomena are under investigation\cite{brisbane,Walt2, 
PRL1,Temmerman-NiO}. The SIC-LSD scheme is a
method of improving the LSD approach by subtracting the spurious interaction of
each occupied electron with itself from the usual LSD approximation to the total
energy. This yields a much-improved description of static Coulomb
correlation effects over the conventional LSD approximation. Examples of the
benefits of this approach have been demonstrated in many applications 
including studies of the Hubbard model,\cite{Rapid,vogl-hub} the 3d 
monoxides,\cite{PRL1,Temmerman-NiO} La${}_2$CuO${}_4$, 
\cite{PRL2,Temmerman-La2} $f$-electron systems,\cite{nature,Leon1,Leon2} 
solid hydrogen,\cite{SSC} orbital ordering,\cite{Rik} and metal-insulator
transitions\cite{Walt1}

In the SIC-LSD method the total energy is minimised with respect to both 
the electron density and the number of electrons from which the
self-interaction has been subtracted. This leads to a determination of the
nominal valence in solids defined as the integer number of electrons 
available to form energy bands
\begin{eqnarray}
N_v=Z-N_{core}-N_{SIC},
\end{eqnarray}
where $N_{core}$ is the number of atomic core electrons, $Z$ is the atomic
number and $N_{SIC}$ is the number of states for which the self-interaction
correction has been removed. This definition means that $N_v$ is 2 for
systems normally thought of as divalent and 3 for systems that are
trivalent, as one might expect.

\section{Results and Discussion}

Most of the rare earth chalcogenides and pnictides crystallise in the 
common rocksalt structure. We have performed SIC-LSD calculations for 
Eu chalcogenides and pnictides using an LMTO method for the band 
structure.\cite{tblmto} Calculations of the electronic structure have 
been performed in the ferromagnetic state for all the materials in 
both the divalent and trivalent state. Whichever of these we find 
with the lower total energy should be the stable valence.

It is well-known that the LSD approximation to density functional theory
does not predict band gaps correctly. The SIC-LSD approach has been shown 
to improve the calculation of band gaps\cite{Temmerman-NiO}, and we would 
expect that this method could reproduce trends in band gaps fairly 
reliably.

\subsection{Europium Chalcogenides}

In figure 1 we show the calculated and experimental lattice constants for 
all the chalcogenides. We also show the energy differences between the 
two valence states. It is clear that all the europium chalcogenides are 
divalent. This is as we would expect on simple shell-filling grounds. We 
can also see from this figure that the calculated lattice constants are 
in good agreement with experiment. The energy difference between the two 
valence states is fairly independent of chalcogenide for S, Se and Te 
and we observe that there is a clear correlation between the lattice 
constant and the difference in energy between the divalent and trivalent 
states.
\begin{figure}
\includegraphics[scale=0.30,angle=270]{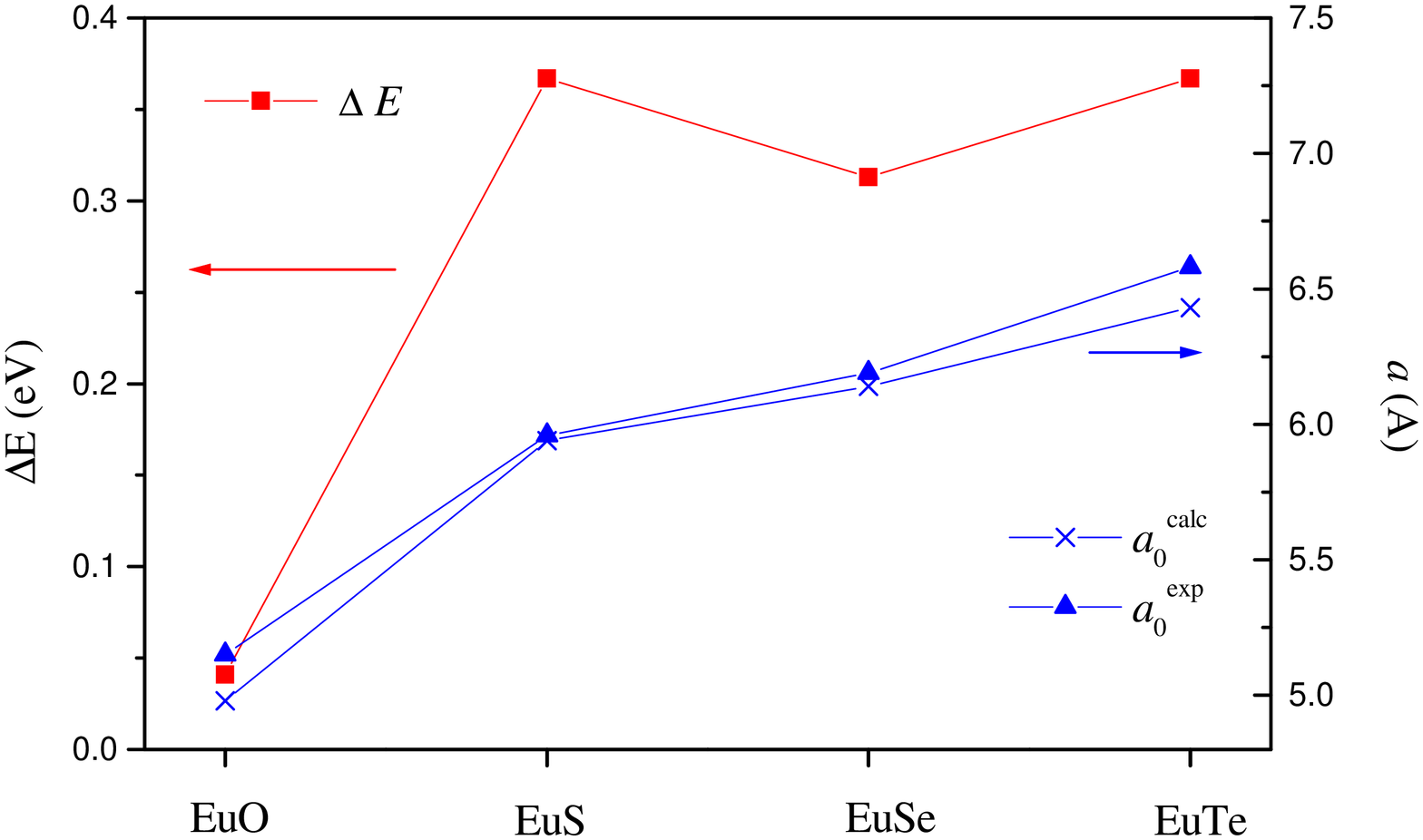}
\caption{The values of the lattice constant for the europium chalcogenides
from experiment\cite{gasg} (triangles) and this calculation ($\times$). Also 
shown is the calculated energy difference between the divalent and trivalent 
form (squares). A positive value means divalency while a negative value
indicates trivalency. For consistency with previous results the energies
here include the 43mRy calibration determined for the elemental rare earths 
and their sulphides by Strange {\it et. al.}\cite{nature}}
\label{fig1}
\end{figure}
\begin{figure*}
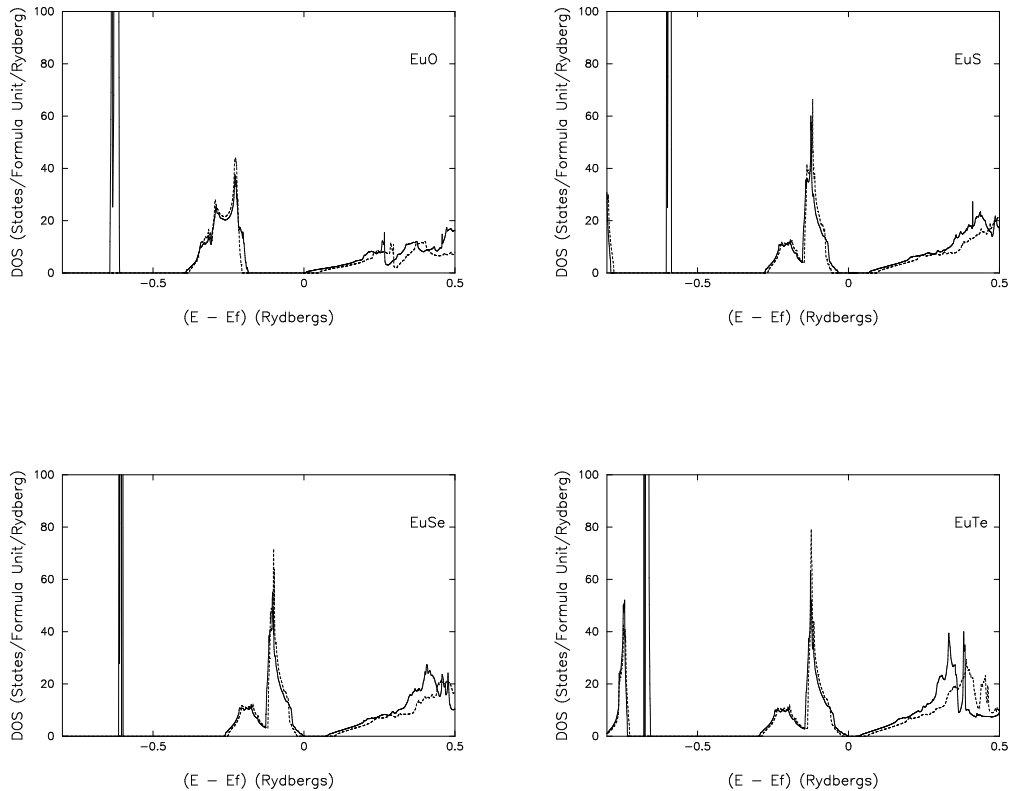

\begin{tabular}{l c c c c c c r}
\includegraphics[scale=0.29,angle=0]{EUOSP.ps} & & & \hspace{2truecm} & & & & 
\includegraphics[scale=0.29,angle=0]{EUSSP.ps} \\
\includegraphics[scale=0.29,angle=0]{EUSESP.ps} & & & \hspace{2truecm} & & & &
\includegraphics[scale=0.29,angle=0]{EUTESP.ps} \\
\end{tabular}
\label{fig2}
\vskip 5mm
\caption{The spin-resolved density of states for the divalent europium 
chalcogenides (majority spin (full line) and minority spin (dashed line)). 
In EuO the calculated energy gap is 2.5 eV compared with the experimental 
value of 1.2 eV.}
\end{figure*}

In figure 2 we show the density of states for divalent EuO, EuS, EuSe and
EuTe. At low energies around -0.6 to -0.7 Rydbergs are the seven fully 
occupied $f$-states in the majority spin channel. Above these is a gap to 
states which are predominantly chalcogenide $p$-bands. Then there is a 
further gap around the Fermi energy. Above $E_f$ are the Eu $s-d$ bands
which include some hybridised chalcogenide $p$-character. The unoccupied 
$f$-states are well above the Fermi energy and not much of them can be 
seen on these figures. There is a large increase in lattice constant as 
we proceed down the column in the Periodic Table with a corresponding 
sharpening of the features of the density of states. This is particularly 
clear in the chalcogenide $p$-states. In figure 2 we have decomposed the 
density of states by spin. The $f$-bands define the majority spin of 
course. However we can immediately observe that there is interesting 
behaviour closer to the Fermi energy. The chalcogenide $p$-states couple 
antiferromagnetically to the rare earth sites, presumably via a 
super-exchange mechanism, as can be observed in the magnetic moments. 
For all the chalcogenides we find the magnetic moment on the rare earth 
site between $7.05\mu_B$ and $7.08\mu_B$ and the chalcogenide moment 
is between  $-0.05\mu_B$ and $-0.08\mu_B$. The occupied majority spin 
states extend up to a higher energy than the occupied minority spin 
states. On the other hand the majority spin conduction band states 
extend down to a lower energy than the minority spin conduction band 
states. This means that there is a significantly lower band gap for the 
majority spin than for the minority spin electrons. In Table I we show 
the calculated band gap for all the chalcogenides decomposed by spin and 
the calculated exchange splitting of the bottom of the conduction band.
\begin{table}
\begin{ruledtabular}
\caption{The first row is the band gaps for the europium chalcogenides. 
It is clear that the gap for the majority spin electrons is significantly 
smaller than the gap for the minority spin electrons. The second row is the
spin splitting ${\Delta E}_{ex}$ at the bottom of the conduction band. All energies 
are in eV.}
\begin{tabular}{ccccccccc}
 & \multicolumn{2}{c}{EuO} & \multicolumn{2}{c}{EuS} & \multicolumn{2}{c}{EuSe} 
& \multicolumn{2}{c}{EuTe} \\
 & $\uparrow$ & $\downarrow$ & $\uparrow$ & $\downarrow$ 
           & $\uparrow$ & $\downarrow$ & $\uparrow$ & $\downarrow$ \\
gap & 2.5 & 3.4 & 1.3 & 2.0 & 1.0 & 1.3 & 0.4 & 1.1 \\  
 ${\Delta E}_{ex}$ & \multicolumn{2}{c}{0.62} & \multicolumn{2}{c}{0.39} &
\multicolumn{2}{c}{0.38} & \multicolumn{2}{c}{0.39} \\
\end{tabular}
\end{ruledtabular}
\vskip 2mm
\end{table}
These results unambiguously confirm the recent experiments by Steeneken 
{\it et al.} which showed that EuO is a small band gap semiconductor and that 
in the ferromagnetic state the charge carriers are almost entirely in one spin 
direction. Our calculations suggest that there may be a range of temperatures 
for which this is true for all the europium chalcogenides in their 
ferromagnetic state. Furthermore these results have a clear implication that 
the band gaps can be controlled by a judicious choice of material and doping.
Steeneken {\it et al.} estimate the exchange splitting in the
conduction band of EuO as 0.6 eV, and again we find excellent agreement with
this value. 
\begin{figure}
\includegraphics[scale=0.30,angle=270]{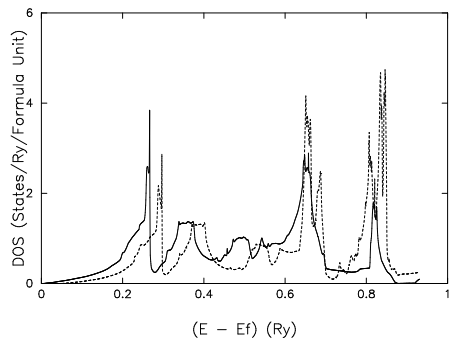}
\caption{The unoccupied region of the oxygen $p$-resolved density of states for 
EuO. The full line is for majority spin and the dashed line for minority spin. 
The Fermi energy is zero on this scale. The positions of the peaks compare very 
favourably with the XAS measurements of Steeneken {\it et al.}\cite{steen}.}
\label{fig3}
\end{figure} 

Steeneken {\it et al.} have also determined the X-ray absorption
spectrum for EuO at the oxygen K-edge. They have shown that it is in good
agreement with the density of states calculated with the LDA + U method
with $U=7.0$~eV. In figure 3 we show the spin-resolved unoccupied $p$-density 
of states on the oxygen site calculated with the SIC-LSD method which is
completely first principles and comparison with the XAS spectrum shows that
it also exhibits good agreement with the XAS results. In the experiment
the principal peaks occur at 532.5 eV with a shoulder at 534.5eV and a 
double peak structure at 536.0 and 538.0 eV. These features match well 
with our density of states peaks at 0.25~Ry (3.4~eV), 0.37~Ry (5.0~eV), 
0.65~Ry (8.8~eV) and 0.82~Ry (11.2~eV). 

\subsection{Europium pnictides}
\begin{figure}
\includegraphics[scale=0.30,angle=270]{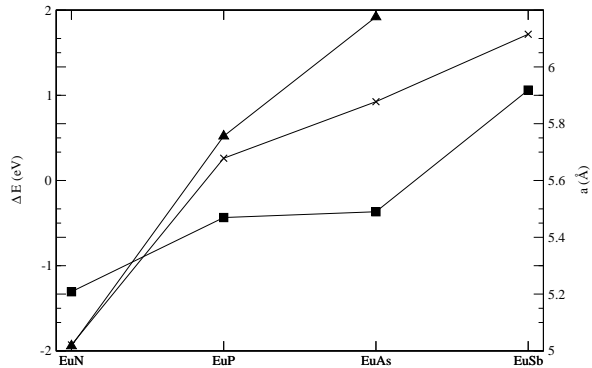}
\caption{The values of the lattice constant for the europium pnictides
on the NaCl structure from the SIC-LSD calculation. $\times$ represent 
the calculated values and triangles are the experimental 
values\cite{Pears}. The large discrepancy for EuAs is because it is 
found to be just trivalent on the NaCl structure whereas on the true 
crystal structure it is divalent (see text). The calculated divalent 
lattice constant is 6.076$A^o$ on the NaCl structure. We have been 
unable to find an experimental value for EuSb. Also shown is the 
calculated energy difference between the divalent and trivalent forms 
(squares). For consistency with previous results the energies here 
include the 43mRy calibration introduced by Strange {\it et 
al.}\cite{nature}}
\label{fig4}
\end{figure}

In figure 4 we show the equilibrium lattice constant for the europium 
pnictides. We also show the energy difference between the divalent and
trivalent configurations (a positive value of $\Delta E$ indicates divalency). 
It is clear that EuN and EuP are trivalent and EuSb is divalent. Although 
the calculation predicts that EuAs is trivalent the energy difference is 
small and is within the margin of uncertainty for the calculation, possibly 
indicating intermediate valence in this compound. For these materials the 
energy difference between the two valence states is strongly dependent on 
the pnictide concerned.
\begin{figure*}
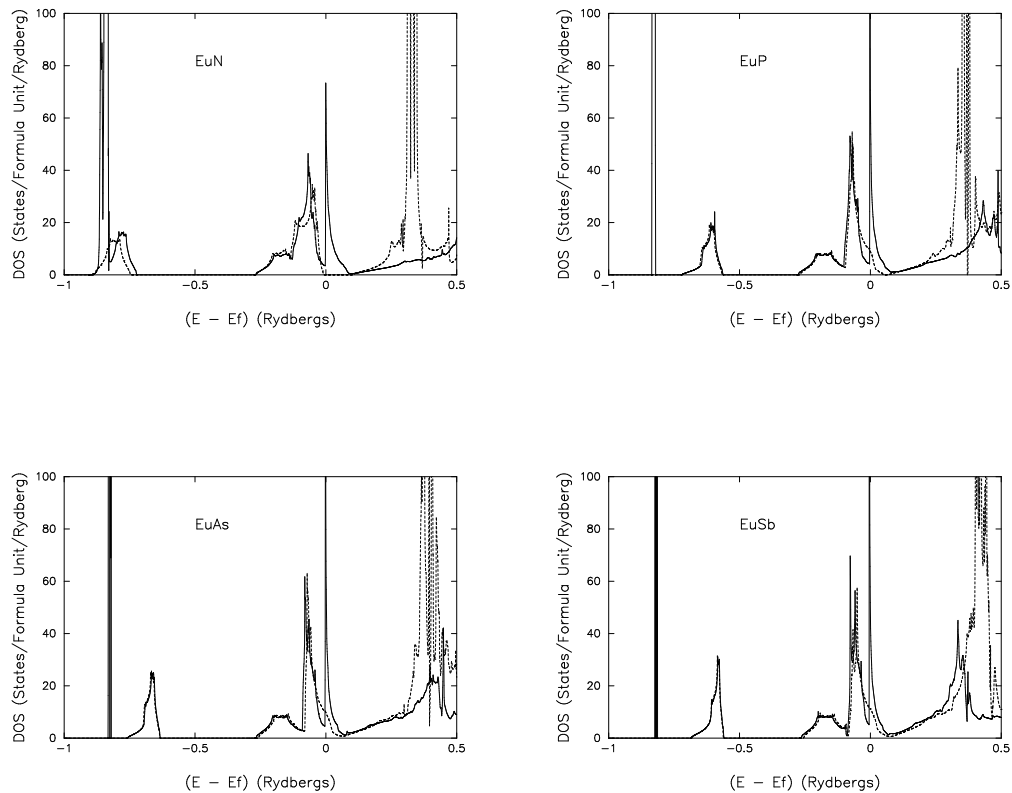

\begin{tabular}{l c c c c c c r} 
\includegraphics[scale=0.29,angle=0]{EUNSP.ps} & & & \hspace{2truecm} & & & & 
\includegraphics[scale=0.29,angle=0]{EUPSP.ps} \\
\includegraphics[scale=0.29,angle=0]{EUASSP.ps} & & & \hspace{2truecm} & & & & 
\includegraphics[scale=0.29,angle=0]{EUSBSP.ps} \\
\end{tabular}
\label{fig5}
\vskip 5mm
\caption{The spin-resolved density of states for Europium Pnictides,
in the trivalent state (majority spin = full line, minority spin = 
dashed line).}
\end{figure*}

These results are in full agreement with the assertion of 
Hulliger\cite{hull} who states that the EuN and EuP are known to be 
trivalent, while EuAs is known to contain some divalent ions. We are 
not aware of any definitive measurement of the valence state of EuSb, 
but clearly if the trend continues it will be divalent. In reality 
EuAs does not have the rocksalt crystal structure and we have been 
unable to discover the true crystal structure of EuSb. EuAs adopts the 
Na$_2$O$_2$ crystal structure which is a distortion of the NiAs 
structure due to the formation of anion-anion pairs.\cite{hull} We 
speculate that the appearance of the divalent ions in the material is 
responsible for this change in crystal structure. Calculations on the
experimentally observed crystal structure support this viewpoint because 
we find that the divalent state of EuAs is favoured by about 0.8~eV per 
formula unit in the Na$_2$O$_2$ crystal structure compared with the 
trivalent state being favoured by 0.2~eV in the NaCl structure. We also 
note from figure 4 that the correlation between lattice constant and the 
energy difference between the divalent and trivalent states is less 
pronounced for the pnictides than for the chalcogenides. This energy 
difference between divalent and trivalent states varies more in the 
pnictides than in the chalcogenides. Note that the scales of Figs. 
\ref{fig1} and \ref{fig4} run over 0.4~eV and 2.0~eV respectively. These 
results are qualitatively similar to those obtained for the ytterbium 
pnictides\cite{yb1,yb2} where there is also an increasing tendency 
towards divalency as one goes down the pnictide column of the Periodic 
Table. However, in that case the divalent state is not reached and 
Yb has a tendency to form highly enhanced heavy fermion systems 
such as YbBiPt for example.
\begin{table}
\begin{ruledtabular}
\caption{The first row is the spin contribution to the magnetic moment for the 
divalent europium pnictides. The second row is the spin magnetic moment for the 
trivalent europium pnictides. All magnetic moments are in Bohr units. The
magnetic moments are further decomposed by site.}
\begin{tabular}{ccccccccc}
 & \multicolumn{2}{c}{EuN} & \multicolumn{2}{c}{EuP} & \multicolumn{2}{c}{EuAs} 
& \multicolumn{2}{c}{EuSb} \\
Eu(2+) & 6.98 & -0.98 & 7.03 & -0.13 & 7.05 & -0.04 & 7.06 & -0.05 \\
Eu(3+) & 6.29 & -0.30 & 6.48 & -0.18 & 6.54 & -0.18 & 6.67 &-0.12 \\
\end{tabular}
\end{ruledtabular}
\vskip 2mm
\end{table}

In Table II we show the magnetic moments of the europium pnictides in both
the divalent and trivalent forms. In the trivalent form the behaviour is as
would be expected. The moments on each site decrease in magnitude as the
atoms get closer together. The pnictide moment has the opposite direction
to the europium moment. In the divalent state EuP, EuAs and EuSb all have
europium moments close to 7$\mu_B$ and the induced moment on the pnictide
site is fairly small. This is not the case for EuN where the lattice constant
is around 15$\%$ smaller than in the other pnictides. In EuN there is
stronger overlap of the nitrogen $p$-electrons and the Eu $f$-electrons
leading to the nitrogen having a magnetic moment close to 1$\mu_B$ in the
opposite direction to the Eu moment. Comparison of the electronic structure
of the other divalent pnictides with EuN shows that in EuN there has been a
transfer of weight of around 0.7 electrons from the rare earth $s-d$ bands
to the minority spin nitrogen $p$-band. 

In figure 5 we show the densities of states for the europium pnictides in
the trivalent state. These exhibit some similarity to, and some key 
differences from, 
the densities of states of divalent chalcogenides. The
%the equivalent pictures for the chalcogenides. The 
pnictide semi-core $s$-bands are above the localised europium 4f levels in 
the pnictides while they are below the $f$-levels in the chalcogenides. 
However, the dramatic differences occur around the Fermi energy. The 
pnictides only have three $p$-electrons and so three electrons from the 
europium are required to fill this electron shell. This is done by the two 
$5d-6s$ electrons and one europium $f$-electron. Trivalency leads to a 
single majority spin $f$-state that is virtually empty sitting very close to the 
Fermi energy. This is the energetically favoured state certainly for EuN 
and EuP and so they are definitely trivalent. As we proceed down the 
pnictide column of the Periodic Table this $f$-level becomes progressively 
more occupied, signalling the change of valence as discussed by Strange 
{\it et al.}\cite{nature} In the pnictides the empty minority spin $f$-states 
can be seen clearly about 0.3-0.4 Rydbergs above E$_{f}$. In all the pnictides
the empty majority spin $f$-state overlaps with the pnictide $p$-band and
pins the Fermi energy. Hence there are some heavy $f$-electrons at
$E_f$ and also a few holes in the pnictide $p$-band. As we go down the
pnictide column  the Fermi energy penetrates more and more into the narrow 
$f$-peak at $E_f$ (this effect is too small to be clearly visible in figure
5). This means that there are more occupied band-like $f$-states and 
according to the suggestion of Strange {\it et al.}\cite{nature} this 
implies a greater tendency to divalency, and this is indeed what we observe 
in the energy difference between the two valencies. It is also worth noting 
that the densities of states shown in figure 5 bear a remarkable similarity 
to those shown by Svane {\it et al.}\cite{yb2} for the ytterbium pnictides. 
This is not surprising given that both classes of materials have a single 
$f$-band very close to the Fermi energy.
\begin{figure}
\begin{tabular}{l c r}
\includegraphics[scale=0.31,angle=0]{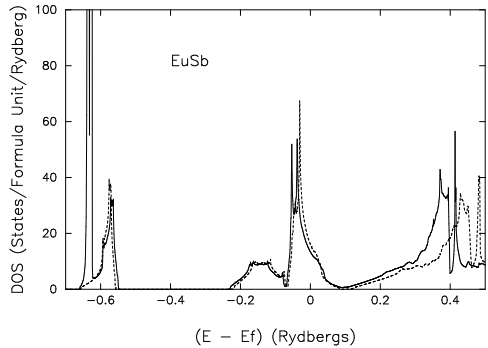} & & \\
\end{tabular}
\caption{The spin-resolved density of states for Europium Antimonide
in the divalent state (majority spin = full line,  minority spin = dashed 
line).}
\label{fig6}
\end{figure}

The density of states again yields very interesting behaviour for these
materials when we decompose it by spin. For EuN we note from figure 5 that
the minority spin density of states is zero at the Fermi energy. The pnictide
$p$-band density of states falls to zero just below the Fermi energy. The 
empty $f$-state close to the Fermi energy is a pure majority spin state. It 
hybridises with majority spin $p$-states and draws some pnictide majority spin
$p$-character above the Fermi energy. There is no minority spin character at the 
Fermi energy at all. So the calculation predicts that EuN is a half metallic 
ferromagnet with a minority spin band gap of $1.4$~eV and a z-component of 
spin-magnetic moment of $6\mu_B$. If we include the expected orbital moment 
according to Hund's rules this would give a total moment of $3 \mu_B$. As we 
proceed down the pnictide column of the Periodic Table there is a transfer of 
around 1.4 electrons from the pnictide $p$-states to the Eu $s-d$ bands 
and the $f$-state which sits close to the Fermi energy. The bulk of this 
transfer occurs between EuN and EuP, although it continues as we go down the 
Periodic Table, and it results in the other europium pnictides being normal 
metals. To understand the transition from half-metallic to full metallic 
behaviour we need to make a detailed study of the density of states. As we go 
from EuN to EuP, the lattice constant increases, the minority spin bands move 
slightly up in energy. The minority spin pnictide 
$p$-states then become less filled and the majority spin Eu $f$-band at the 
Fermi energy and some of the $s-d$ states have a higher occupancy. During this 
transition the minority spin $p$-bands actually rise above the Fermi energy 
and the gap is destroyed in this spin channel. Thus, EuN is a half-metallic 
metal, while EuP and the other europium pnictides are normal metals. This 
extreme sensitivity of the bands to changes in the lattice constant 
also accounts for the strong dependence of the energy difference between the 
two valence states on the pnictide atom.

In figure 6 we show the density of states for EuSb in the divalent state.
This is the calculated ground state. We note that the occupied $f$-states have
risen considerably from their energy in the trivalent state. The results
appear similar to those for the europium chalcogenides (although there is
one fewer electron in the antimonide than in the telluride). In the divalent
state EuSb is clearly a normal metal. About 0.1 Rydbergs above E$_{f}$, 
EuSb just fails to open up a gap which would correspond to the energy gaps 
in the chalcogenides.

\vskip 5mm
\section{Conclusions}

In this paper we have reported a series of electronic structure calculations 
for the europium chalcogenides and pnictides. These calculations tell us 
several things that are interesting from both a fundamental and applied point 
of view. Firstly we have shown that the Eu chalcogenides in their 
ferromagnetic state are semiconductors and the band gaps for the different
spin channels are very different. Therefore we have materials whose carriers 
will be more or less 100\% spin-polarised in the ambient temperature range.
We suggest that, by alloying, it should be possible to create materials
with a range of differential band gaps for the different spins. Secondly, 
we have shown that EuN is a half-metallic system with a substantial moment in 
its ferromagnetic state while the other europium pnictides are normal metals. 
Again we might suggest that alloying of the ferromagnetic pnictides would 
allow us to have a half metallic system with a range of possible energy gaps 
in the minority spin channel. Thirdly our calculations strongly suggest that 
the position of the occupied $f$-levels in europium materials is dependent on 
the valence, but not on the chemical environment, of the europium ion.
Finally we predict a valence transition in the europium pnictides as we
proceed down the pnictide column of the periodic table which may account for
the observed deviation from the common rocksalt crystal structure expected
for EuAs and EuSb. It seems that the europium chalcogenides and pnictides are 
possible candidates for spintronic and spin filtering applications.

\end{document}